%% file: qsd09.tex
\documentclass[final,3p,times,fleqn,sort&compress]{elsarticle}

\usepackage{cleveref}
\usepackage{geometry}
\usepackage{pifont}
\usepackage{graphics}
\usepackage{fleqn}
\usepackage{amsmath}
\usepackage{amsfonts}
\usepackage{dsfont}
\usepackage{color}
\usepackage{dcolumn}
\usepackage[cp1251]{inputenc}
\usepackage[english]{babel}
\usepackage{epsfig}
\usepackage{array}
\usepackage{comment}
\usepackage[12hr]{datetime}
\input{qsd_header}

\includecomment{derivation}
\includecomment{notes}
\def\etal{\textit{et al. }}
\def\htot{H_{\textup{tot}}}
\def\hsys{H_{\textup{sys}}}
\def\hint{H_{\textup{int}}}
\def\heff{H_{\textup{eff}}}

\def\leff{L_{\textup{eff}}}

\def\ok{\omega_k}
\def\adk{a^\dagger_k}
\def\ak{a_k}
\def\ih{\frac{i}{\hb}}

\begin{document}
\begin{frontmatter}
\title{Solving non-Markovian open quantum systems\\with multi-channel reservoir coupling}

\author[ur]{Curtis J. Broadbent\corref{cor}}
\ead{curtis.broadbent@rochester.edu}
\author[sit]{Jun Jing}
\author[sit]{Ting Yu}
\author[ur]{Joseph H. Eberly}
\cortext[cor]{Corresponding author.}

\address[ur]{Rochester Theory Center, and Department of Physics and Astronomy, University of Rochester, Rochester, New York 14627, USA}
\address[sit]{Center for Controlled Quantum Systems, and the Department of Physics and Engineering Physics, Stevens Institute of Technology, Hoboken, New Jersey 07030, USA}

\date{\today}
\begin{abstract}
  We extend the non-Markovian quantum state diffusion (QSD) equation to open quantum systems which exhibit multi-channel coupling to a harmonic oscillator reservoir. Open quantum systems which have multi-channel reservoir coupling are those in which canonical transformation of reservoir modes cannot reduce the number of reservoir operators appearing in the interaction Hamiltonian to one. We show that the non-Markovian QSD equation for multi-channel reservoir coupling can, in some cases, lead to an exact master equation which we derive. We then derive the exact master equation for the three-level system in a vee-type configuration which has multi-channel reservoir coupling and give the analytical solution. Finally, we examine the evolution of the three-level vee-type system with generalized Ornstein-Uhlenbeck reservoir correlations numerically.   
\end{abstract}
\begin{keyword}
  Open Quantum Systems \sep Decoherence \sep Quantum State Diffusion \sep Master Equation \sep Stochastic \schrodinger Equation \sep Non-Markovian Systems
\end{keyword}
\end{frontmatter}
\section{Introduction}
Master equations have served as an essential tool in the dynamical analyses of open quantum systems since their introduction in a quantum mechanical setting by Pauli in 1928 \cite{Pauli1928}. At their core, master equations simplify the complexity inherent in the dynamics of a large reservoir by recording only the influence of the reservoir on the evolution of the system. Master equations also have the attractive feature that their solution automatically gives the evolution of any system observable. In contrast, while the Heisenberg picture approach may often be less complicated, the results are usually applicable to only one or two Heisenberg operators \cite{Wodkiewicz1976}. When the coupling to the reservoir is weak and when the correlation times of the reservoir are negligible compared to the evolution time of the system, the coupling may be said to be Markovian, and the so-called Lindblad master equation may be used to approximate the system evolution \cite{Gorini1976,Lindblad1976}. In cases where the coupling to the reservoir is non-Markovian, alternative techniques must be used to determine the system dynamics.

Methods for determining the dynamics of an open quantum system with non-Markovian coupling to a reservoir include the quantum trajectory approach. Quantum trajectory methods use a stochastic \schrodinger equation to evolve pure system states subject to a stochastic process. The evolution of the system density operator is given by the mean of the solution to the stochastic equation. The average of repeated numerical implementations of the stochastic evolution equation can be used to determine the evolution of the system when a formal solution to the stochastic equation cannot be found. 

A primary example of a quantum trajectories approach is the non-Markovian quantum state diffusion (QSD) equation. First introduced by Di\'osi and Strunz \cite{Diosi1997}, the non-Markovian quantum state diffusion equation is a generalization of the Markovian QSD equation \cite{Gisin1992,Gisin1993,Gisin1993a}. In contrast to the Markovian QSD equation which uses a white noise process in the stochastic evolution equation, the non-Markovian QSD equation is built around a colored noise process. The non-Markovian QSD equation has been used to examine a wide variety of non-Markovian open quantum systems, including the spin-N system, the N-cavity model, the N-qubit model, and many others \cite{Diosi1998,Yu1999,Strunz1999,Yu2004,Strunz2004,Jing2010,Jing2010a,Jing2011}. The non-Markovian QSD equation has also been used as an analytical tool, since, in some cases, the non-Markovian QSD equation can be used to derive an exact master equation. For instance, the exact master equation for quantum Brownian motion \cite{Hu1992,Halliwell1996,Ford2001} has also been derived using the non-Markovian QSD equation \cite{Strunz2004}. The non-Markovian quantum state diffusion equation is exact and applies to open quantum systems in  which, a) the initial state of the total system is factorizable into an initial system state and a thermal reservoir state, and b) the system couples to the reservoir though a single channel. 

Single-channel reservoir coupling occurs when the reservoir operators entering into the interaction Hamiltonian are mutually proportional. Consider, for example, the following interaction Hamiltonian, 
\begin{equation}\label{hint}
  \hint=\sum_{m}\left(L_m\otimes B^\dagger_m+\hc\right),
\end{equation}
where $L_m$ are operators of the system, $B_m=\sum_kg_{mk}a_k$ are the reservoir operators expressed in terms of $a_k$, the annihilation operators for mode $k$ of a boson reservoir, and $g_{mk}$ is the coupling constant between system operator $L_m$ and reservoir mode $k$. If $B_m=\kappa_mB$ for all $m$, then the reservoir operators are mutually proportional and the interaction Hamiltonian may be reduced to interactions between a single reservoir operator, $B$, and an effective system operator $\leff=\sum_m\kappa^*_mL_m$,
\begin{align}
  \hint=\leff\otimes B^\dagger+\hc\label{singlemodehint}
\end{align}
In the context of open quantum systems where the system consists of multiple particles which are coupled to a common reservoir, interactions of the form of \eqref{singlemodehint} are often described as having collective decoherence and are an essential part of the theory of decoherence free subspaces \cite{Zanardi1998,Lidar2003}. 

The standard non-Markovian quantum state diffusion equation may be used for interaction Hamiltonians which take the form of \eqref{singlemodehint}. Such interactions may be called single-channel reservoir coupling interactions because a canonical transformation can always be made to a new set of reservoir modes
\begin{align}
  b_j=\sum_ku_{jk}a_k,
\end{align}
for which $b_0=\eta B$, where $\eta=1/\sqrt{\sum_k|g_k|^2}$. Consequently, we have $\hint=(\tilde{L}_{\textup{eff}}\otimes b^\dagger_0+\hc)$ where $\tilde{L}_{\textup{eff}}=\leff/\eta$. In contrast, multi-channel reservoir coupling occurs when a canonical transformation of reservoir modes cannot reduce the number of reservoir operators appearing in the interaction Hamiltonian to one (up to the Hermitian conjugate operation). Alternatively, we may say that the interaction Hamiltonian exhibits multi-channel reservoir coupling if the reservoir operators are not mutually proportional.

Apart from the fundamental interest in solving open quantum systems with general multi-channel coupling, there is one system currently under both theoretical and experimental investigation which can exhibit multi-channel reservoir coupling. Donor-based charge quantum bits in a semiconductor host have single-channel reservoir coupling only if the Bohr radii of the s-wave orbitals are exactly equal for all donors \cite{Lastra2011a}. Donor-based charge quantum bits are of particular interest due to their potential use within a scalable architecture for a quantum computer. Since multi-channel reservoir coupling can eliminate decoherence free subspaces which have been proposed as a strategy for avoiding errors in quantum computing \cite{Zanardi1997,Lidar1998,Lidar2003}, studying decoherence in the presence of multi-channel reservoir coupling is of particular importance for this system. 

In this article we will extend the quantum state diffusion method to multi-channel reservoir couplings. As is the case for the standard non-Markovian QSD equation, we will show that an exact master equation may sometimes be derived from the non-Markovian QSD equation for multi-channel reservoir coupling. We will then derive the exact master equation and give the analytical solution for the three-level atom in a vee-type configuration. Finally,  we will provide a toy model for the noise correlations which will be used to examine the way in which multi-channel reservoir coupling can affect system evolution. 

\section{Non-Markovian QSD equation for multi-channel reservoir coupling}
We first derive the non-Markovian QSD equation associated with the interaction Hamiltonian \eqref{hint}. We will take an approach similar to the derivation of Strunz and Yu \cite{Strunz2004}, and consider the case where the reservoir is initially in the vacuum state. A Bogoliubov transformation can be used to transform the thermal-state case to the vacuum-state case \cite{Yu2004}. Adding the system and reservoir Hamiltonians to \eqref{hint} we have,
\begin{equation}
  \htot=\hsys+\hint+\sum_k\hb\ok\adk\ak.
\end{equation}
We go to the interaction picture of the reservoir, and write the \schrodinger equation for the total system, 
\begin{equation}
  i\hbar\partial_t\ket{\Psi_t}=\left[\hsys+\hbar\sum_{mk}\left(g^*_{mk}L_m\adk e^{i\ok t}+g_{mk}L^\dagger_m\ak e^{-i\ok t}\right)\right]\ket{\Psi_t},
\end{equation}
where we use the abbreviation, $\ket{\Psi_t}\equiv\ket{\Psi(t)}$.
We then take the component of the reservoir in the Bargmann state, $\braket{z}{\Psi_t}$, where $\ket{z}\equiv\ket{\{z_k\}}=\otimes_k\hat{z}_k\ket{\textup{VAC}}$ and $\hat{z}_k=e^{z_k\adk}$, 
\begin{equation}\label{stochastic}
  \partial_t\psi_t=-\frac{i}{\hb}\hsys\psi_t+\sum_m\left(z_{mt}^*L_m-L_m^\dagger\int_0^t\!\!ds\,\sum_n\alpha_{mn}(t,s)\frac{\delta}{\delta z_{ns}^*}\right)\psi_t.
\end{equation}
In \eqref{stochastic}, we have defined $z_{mt}^*\equiv-i\sum_kg^*_{mk}z_k^*e^{i\ok t}$,
\begin{equation}\label{correlation}
  \alpha_{mn}(t,s)\equiv\sum_kg_{mk}g^*_{nk}e^{-i\ok (t-s)},
\end{equation}
and the unnormalized system state $\ket{\psi_t(z^*)}\equiv\braket{z}{\Psi_t}$ with shorthand $\psi_t(z^*)=\ket{\psi_t(z^*)}$. In $\psi_t(z^*)$ and in the following, $z^*$ ($z$) is used as shorthand for the vector of complex Bargmann coefficients $\{z^*_k\}$ ($\{z_k\}$). The functional derivative in \eqref{stochastic} arises from applying the chain rule to the derivative of the Bargmann state,
\begin{equation}
  \bra{z}\ak=\frac{\partial}{\partial z^*_k}\bra{z}=\int_0^t\!\!ds\sum_n\frac{\partial z_{ns}^*}{\partial z_{k}^*}\frac{\delta}{\delta z_{ns}^*}\bra{z}.
\end{equation}

  As written, \eqref{stochastic} is the exact evolution equation for $\psi_t(z^*)$ and does not require interpretation as a stochastic equation, i.e. \!given an initial state of the total system $\ket{\Psi_0}$, and a Bargmann state $\ket{z}$, \eqref{stochastic} simply gives the evolution of the Bargmann component of the total system, $\ket{\psi_t(z^*)}$. Since the Bargmann state is an over-complete basis, it can be used to evaluate the trace over the reservoir, giving the reduced density operator of the system,   
  \begin{align}
    \rho_t=\int d^2z\,p(z)\braket{z}{\Psi_t}\braket{\Psi_t}{z}\label{rhoformal},
      \end{align}
  where $d^2z=d^2z_1d^2z_2\cdots$ and 
  \begin{align}
    p(z)=\prod_k\left(\frac{e^{-|z_k|^2}}{\pi}\right).\label{distribution}
  \end{align}
  The normalizing function $p(z)$ appearing in \eqref{rhoformal} is of the precise form of a distribution of independent complex Gaussian random variables. We can therefore consider $z$ to be a multivariate complex Gaussian random variable, where we define the statistical mean $\mathcal{M}\{\cdot\}$ of a stochastic system operator $\Phi(z)$,
  \begin{align}
    \mathcal{M}\{\Phi(z)\}=\int\!\!d^2z\,p(z)\Phi(z).
  \end{align}

  In this context we interpret $\{z_{mt}^*\}$ as a set of continuous Gaussian random processes with correlation functions $\alpha_{mn}(t,s)=\mathcal{M}\{z_{mt}z^*_{ns}\}$, already worked out in \eqref{correlation}. Since the noise processes all arise from the same reservoir modes, the cross-correlation of the noise terms given in \eqref{correlation} is non-zero when $m\neq n$ (in contrast to the noise terms which arrive in dealing with a thermal bath \cite{Yu2004}). Additionally, we may now interpret \eqref{stochastic} as a stochastic evolution equation for $\ket{\psi_t(z^*)}$, which we now call the (unnormalized) stochastic state vector. Finally, the evolution of the system is found by taking the statistical mean of the outer product of the stochastic state vector \cite{Strunz2004}, 
\begin{equation}
  \rho_t=\mathcal{M}\{\proj{\psi_t(z^*)}\}.\label{density}
\end{equation}

As in the standard non-Markovian QSD derivation we consider solutions to \eqref{stochastic} which satisfy the following condition: The functional derivative in the non-Markovian QSD equation may be written in terms of an operator, commonly called the O-operator, which depends on the Bargmann states,
\begin{equation}
  \frac{\delta\psi_t}{\delta z_{ms}^*}=O_m(t,s,z^*)\psi_t.\label{ansatz}
\end{equation}
where $O_m(s,s,z^*)=L_m$ so that \eqref{stochastic} is able to reproduce the QSD equation in the Markov limit \cite{Gisin1992,Gisin1993,Gisin1993a}. 

By defining
\begin{equation}
  \overline{Q}_m(t,z^*)=\int_0^t\!\!ds\left(\sum_n\alpha_{mn}(t,s)O_n(t,s,z^*)\right),
\end{equation}
we can write the non-Markovian QSD equation under the ansatz,
\begin{equation}
  i\hb\partial\psi_t=\heff(t,z^*)\psi_t,\label{compactlinear}
\end{equation}
where $\heff(t,z^*)$ is the effective stochastic Hamiltonian,
\begin{equation}\label{heffective}
  \heff(t,z^*)=\hsys+i\hb\sum_m\left(z_{mt}^*L_m-L_m^\dagger\overline{Q}_m(t,z^*)\right).
\end{equation}

To be self-consistent, the solution condition specified in \eqref{ansatz} must satisfy the following conditions,
\begin{equation}
  \partial_t\frac{\delta\psi_t}{\delta z_{ms}^*}=\frac{\delta}{\delta z_{ms}^*}\partial_t\psi_t,
\end{equation}
giving rise to the consistency equations for the O-operator $O_m(t,s,z^*)$,
\begin{equation}
  \partial_tO_m=-\frac{i}{\hb}\left[\heff(t,z^*),O_m\right]-\sum_nL_n^\dagger\frac{\delta}{\delta z_{ms}^*}\overline{Q}_n(t,z^*).\label{Oconsistency}
\end{equation}

Finally, by making a Girsanov transformation \cite{Gatarek1991,Yu1999}, we can write the nonlinear non-Markovian QSD equation, which is more suitable for numerical simulation,
\begin{align}
\partial_t\tilde{\psi}_t=-\frac{i}{\hb}\hsys\tilde{\psi}_t+\sum_m\Big[\tilde{z}^*_{mt}\Delta(L_m)-\Delta\left(\Delta(L_m^\dagger)\overline{Q}_m(t,\tilde{z}^*)\right)\Big]\tilde{\psi}_t,
\label{nonlinear}
\end{align}
where $\Delta(A)=A-\langle A \rangle$ for any operator $A$ and $\tilde{z}^*_{mt}=z^*_{mt}+\sum_n\int_0^t\alpha_{mn}(t,s)\langle L_m^\dagger\rangle ds$.

Equations (\ref{compactlinear}-\ref{heffective}) and (\ref{Oconsistency}-\ref{nonlinear}) are the main results of this article. They allow systems with multi-channel non-Markovian reservoir coupling to be evaluated numerically using finite-difference methods. Additionally, in some cases, the quantum state diffusion equations allow for more efficient simulation as compared to master equations in cases where the master equation may also be derived \cite{Diosi1998}. 

When the O-operators are noise-independent, a master equation may be derived, as is also the case for single-channel coupling. Following the treatment of \cite{Strunz2004} we evaluate the time-derivative of the density operator in \eqref{density} with the help of \eqref{compactlinear} to arrive at
\begin{align}
  \frac{d}{dt}\rho_t=-\frac{i}{\hb}[\hsys,\rho_t]+\sum_m\bigg(\Big[\mathcal{M}\big\{\overline{Q}_m(t,z^*)P_t\big\},L^\dagger_m\Big]+\hc\bigg),
\end{align}
where $P_t=\proj{\psi_t(z^*)}$ and we have used the Novikov theorem and the operator ansatz \eqref{ansatz} \cite{Novikov1965,Strunz2004} to write
\begin{align}
  \mathcal{M}\{P_tz_{mt}\}=\mathcal{M}\{\overline{Q}_m(t,z^*)P_t\}.\label{Novikov}
\end{align}
When the O-operators are noise independent, then $\overline{Q}_m(t,z^*)=\overline{Q}_m(t)$ and we can write an exact master equation, 
\begin{align}
  \frac{d}{dt}\rho_t=-\frac{i}{\hb}[\hsys,\rho_t]+\sum_m\bigg(\Big[\overline{Q}_m(t)\rho_t,L^\dagger_m\Big]+\left[L_m,\rho_t\overline{Q}^\dagger_m(t)\right]\bigg),
\end{align}
which is the multi-channel analog of the exact master equation found by Strunz and Yu \cite{Strunz2004} for single-channel reservoir couplings.

\section{Three-level vee-type system}
\begin{figure} 
  \centering
  \includegraphics[scale=.5]{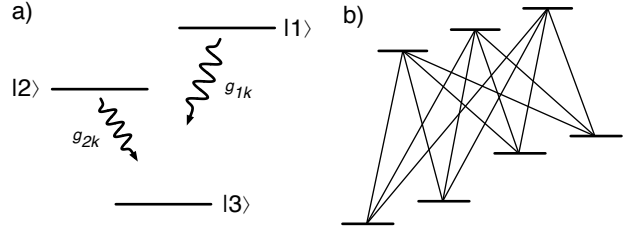}
  \caption{a) Three-level system in vee-type configuration. b) Multi-level vee-lambda system with 3 upper levels and 4 lower levels. The lines connecting levels indicate possible transitions.}
\label{leveldiagrams} 
\end{figure}

We now apply the previous results to demonstrate the utility of the QSD approach for open quantum systems with multi-channel reservoir coupling. We consider the three-level system in a vee-type configuration with multi-channel reservoir coupling, as shown in Fig.~\ref{leveldiagrams}a). We consider the case where the decoherence is generated by the lowering operators, $L_m=\ket{3}\bra{m}$,
\begin{align}
  \hint=\ket{3}\bra{1}\otimes B^\dagger_1+\ket{3}\bra{2}\otimes B^\dagger_2+\hc,
\end{align}
where the excited levels $\ket{1}$ and $\ket{2}$ are separated from the ground state $\ket{3}$ by energies $\hbar\omega_1$ and $\hbar\omega_2$, respectively, so that $\hsys=\hb\omega_1\proj{1}+\hb\omega_2\proj{2}$. We note that in the following, summations will always be over indices corresponding to the two upper levels, $1$ and $2$. When we expand the O-operator as
\begin{align}\label{vooperator}
  O_q(t,s,z^*)=\sum_pf_{qp}(t,s)\ket{3}\bra{p},
\end{align}
we find that the consistency equations reduce to
\begin{align}
  \partial_tf_{qp}=i\omega_pf_{qp}+\sum_mf_{qm}F_{mp}\label{veeconsistency},
\end{align}
where $f_{qp}(s,s)=\delta_{qp}$, and we have defined
\begin{align}
  F_{mp}(t)=\int_0^t\!\!ds\sum_n\alpha_{mn}(t,s)f_{np}(t,s),
\end{align}
so that $\overline{Q}_m(t,z^*)=\sum_pF_{mp}(t)L_p$.  The resulting master equation is given by 
\begin{align}
  \begin{split}
  \frac{d\rho_t}{dt}=-\ih[\hsys,\rho_t]+\sum_{mp}\left(F_{mp}(t)[L_p\rho_t,L_m^\dagger]+F_{pm}^*(t)[L_p,\rho_tL_m^\dagger]\right).\label{formalmaster}
\end{split}
\end{align}

With only slight adjustment, the O-operator in \eqref{vooperator} can be generalized successfully to any system which can be divided into upper and lower levels with transitions forbidden within the upper and lower levels, respectively, and in which the decoherence generators are lowering operators between upper and lower levels \cite{Broadbent2011b}. We call such systems \emph{multi-level vee-lambda systems} since they are a generalization of both the vee and lambda systems. Figure~\ref{leveldiagrams}b gives an example of a 7-level vee-lambda system. The resulting master equation will be of the exact form of \eqref{formalmaster}, albeit with the summation running over all upper and lower levels.

Though the master equation in \eqref{formalmaster} appears similar in nature to approximate master equations (such as the Born-Markov or Lindblad master equations), \eqref{formalmaster} is exact. The time-dependence present in the coefficients encapsulates the total effect of the reservoir, and, in particular, any non-Markovian effects.

For the three-level vee-type system we can find an analytic solution to the master equation. We first define the operator
\begin{align}
  F(t)=\sum_m F_{mn}(t)\ket{m}\bra{n},
\end{align}
which allows us to write
\begin{equation}
  \partial_t\rho=-\frac{i}{\hb}[\hsys,\rho]-(F\rho+\rho F^\dagger)+\traceq{F\rho+\rho F^\dagger}\ket{3}\bra{3}.\label{operatorform}
\end{equation}
The solution to \eqref{operatorform} with initial condition $\rho(0)=\rho_0$ is given by
\begin{subequations}\label{analyticsol}
\begin{equation}
  \rho(t)=p(t)\rho_e(t)+\Big(1-p(t)\Big)\proj{3}\label{rho}
\end{equation}
\textup{where}
\begin{align}  
  p(t)=\traceq{\mathcal{O}^\dagger_t\mathcal{O}_t\rho_0},\quad
\end{align}
  \begin{align} \rho_e(t)=\frac{\mathcal{O}_t\rho_0\mathcal{O}^\dagger_t}{\traceq{\mathcal{O}^\dagger_t\mathcal{O}_t\rho_0}},\label{rhoe}
\end{align}
\textup{and}
  \begin{equation}
    \mathcal{O}_t=\exp\left[-\frac{i}{\hb}\hsys t-\int_0^t\!\!ds\,F(s)\right].\label{decayoperator}
\end{equation}
\end{subequations}

Equation \eqref{analyticsol} gives the general solution of a vee-type three-level system coupled to a vacuum harmonic oscillator reservoir under the rotating wave approximation. General properties of vee-type systems can be deduced from the form of \eqref{analyticsol}. Since $\hsys$ and $F(t)$ have support in only the system subspace defined by the excited states, $\mathcal{H}_e=\textup{Span}[\{\ket{1},\ket{2}\}]$, if the system initial state is excited, $\rho_0\in\mathcal{H}_e$, then $\rho_e(t)$ is also an excited state, $\rho_e(t)\in\mathcal{H}_e$. Additionally, if the initial state is pure, $\rho_0=\proj{\psi_0}$, then $\rho_e(t)=\proj{\psi_t}$ is also pure, where $\ket{\psi_t}\propto\mathcal{O}_t\ket{\psi_0}.$ Consequently, we can interpret the solution as follows: at time $t$ an initially pure and initially excited state will have either decayed to the ground state with probability $(1-p(t))$ or, with probability $p(t)$, have undergone unitary evolution within the excited state subspace.

\section{Ornstein-Uhlenbeck Correlations}
We now consider a toy model for the noise correlations which will allow us to examine the effect that non-Markovian and/or multi-channel reservoir coupling has on the system dynamics. The noise model we will investigate is the generalized Ornstein-Uhlenbeck [OU] correlation function, 
\begin{align}
  \alpha_{mn}(\tau)=\frac{\kappa_m^*\kappa_n\gamma_m\gamma_n}{\gamma_m+\gamma_n+i(\Omega_m-\Omega_n)}\left(e^{-(\gamma_m+i\Omega_m)\tau}\theta(\tau)+e^{(\gamma_n-i\Omega_n)\tau}\theta(-\tau)\right),\label{genOU}
\end{align}
where $\tau=t-s$, and $\theta(t)$ is the Heaviside function. The generalized OU correlation function reproduces the standard OU correlation function when $n=m$,
\begin{align}
  \alpha_{mm}(\tau)=|\kappa_m|^2\frac{\gamma_m}{2}e^{-\gamma_m|\tau|}e^{-i\Omega_m\tau}.\label{standardOU}
\end{align}
We define $\Gamma_m=|\kappa_m|^2$ since it corresponds to the decay rate in the standard OU correlation function. $\Omega_m$ is the central frequency of the OU correlation function. We also see that $1/\gamma_m$ is the correlation time of $B_m$; as $\gamma_m\rightarrow\infty$ the standard OU correlation function goes to a delta function $\alpha_{mm}(\tau)\rightarrow\Gamma_m\delta(\tau)$. For this reason a finite value of $\gamma_m$ signals the departure from the Markov limit of the reservoir coupling to level $m$.

The parameters in \eqref{genOU} can also be described in terms of a quasi-Lorentzian model for the coupling coefficient,
\begin{equation}
  g_m(\omega)=\frac{\kappa_m}{\sqrt{2\pi n'(\omega)}}\frac{\gamma_m}{\gamma_m+i(\omega-\Omega_m)},\label{coupling}
\end{equation}
where $n'(\omega)$ is the spectral density of reservoir modes, $\kappa_m$ is a coupling strength (with units of $\sqrt{\textup{Hz}}$), and $\gamma_m,\Omega_m$ (where $\gamma_m,\Omega_m>0$ and have units of $\textup{Hz}$) are the bandwidth and central frequency of the reservoir coupling. By letting the summation over $k$ in \eqref{correlation} go to an integral over all frequencies (positive and negative), the Lorentzian coupling coefficient \eqref{coupling} reproduces the noise correlations in \eqref{genOU}. In the following we will describe the parameters of the OU correlations in terms of their properties in the coupling coefficients.

Using the generalized OU correlation function, the evolution of the time dependent operator $F(s)$ in \eqref{analyticsol} is given by
\begin{equation}
  \partial_tF_{mn}=\alpha_{mn}(0)-(\gamma_m+i(\Omega_m-\omega_n))F_{mn}+\sum_pF_{mp}F_{pn}.
  \label{OUconsistency}
\end{equation}
This set of coupled non-linear first order equations can be solved numerically for particular choices of the six parameters  in the correlation function $\{\kappa_m,\gamma_m,\Omega_m\}$ and the two atomic energies $\{\omega_m\}$. The solution can then be combined with the analytical solution in \eqref{analyticsol} to give the complete dynamics of the atomic density operator.

\section{Single-channel reservoir coupling}
We now investigate some choices for the coupling parameters to show how multi-channel reservoir coupling manifests itself in this system.  To do this we first examine the reduction of the consistency equation to the single-channel coupling case, $B_m=\kappa_mB$, which implies that $\gamma_m=\gamma$ and $\Omega_m=\Omega$. If, in addition, we consider the case where levels $1$ and $2$ are degenerate, $\omega_m=\omega_0$, then we can derive the evolution of the coefficients $F_{mn}(t)$ analytically. By defining $\Delta=(\omega_0-\Omega)$, $\beta=-(\gamma-i\Delta)/2$, and $\eta^2=\kappa^2\gamma/2-\beta^2$ (where $\kappa^2=|\kappa_1|^2+|\kappa_2|^2=\Gamma_1+\Gamma_2$), it can be shown that
\begin{equation}
  F_{mn}(t)=\frac{\kappa_m^*\kappa_n}{\kappa^2}Q(t),
\end{equation}
where $Q(t)$ satisfies the differential equation 
\begin{equation}
  \partial_tQ=(Q+\beta)^2+\eta^2,
  \label{Qeq}
\end{equation} 
with solution
\begin{equation}
  Q(t)=\kappa^2\frac{\gamma}{2}\left(\frac{\sin\eta t}{\eta\cos\eta t-\beta\sin\eta t}\right)\label{Q},
\end{equation} 
under the initial condition $Q(0)=0$. 

$Q(t)$ governs the decoherence behavior for many qualitatively different parameter regimes. For some of these regimes, $Q(t)$ exhibits complex infinities similar to those described by Di\'osi \etal  \cite{Diosi1998}. These infinities pose no difficulty to the evolution of the quantum system as discussed by Strunz \etal \cite{Strunz1999}. That conclusion is confirmed here since \begin{align}
  \exp\left[-\int_0^t\!\!ds\,Q(s)\right]=\frac{e^{\beta t}}{\eta}(\eta\cos\eta t-\beta\sin\eta t),
\end{align}
so that infinities in $Q(t)$ correspond to zeros in the evolution of $\rho_t$.

If we define the state 
\begin{equation}
  \ket{\phi^+}=\frac{1}{\kappa}\sum_m\kappa_m\ket{m},
\end{equation}
we find that we can write the master equation as 
\begin{equation}
  \partial_t\rho_t=-\frac{i}{\hbar}[\hsys,\rho_t]+Q(t)\left([\leff\rho_t,\leff^\dagger]+[\leff,\rho_t\leff^\dagger]\right),\label{mastervee}
\end{equation}
where  $\leff=\ket{3}\bra{\phi^+}$. Written this way it is easy to see that $\ket{\phi^-}\propto\kappa_2\ket{1}-\kappa_1\ket{2}$ is a trivial solution to \eqref{mastervee}. Alternatively, it can be shown that $\ket{\phi^-}$ satisfies the conditions of a decoherence free subspace \cite{Zanardi1997,Zanardi1998,Lidar1998}. Additionally, since any excited state will eventually decay to a mixture of $\ket{\phi_-}$ and the ground state, the vacuum reservoir induces coherence between levels $1$ and $2$, for all values of $\gamma>0$. This effect was first discovered for the three-level vee-type system by Agarwal \cite{Agarwal1974} in the Markov limit and is known as vacuum-induced coherence\footnote{In the literature there appear to be two scenarios related to vacuum-induced coherence. The other one includes classical fields which participate in the effect. See, for instance Ref.~\cite{Berman1998}.}. It should be noted that the three-level vee-type system under a rotating wave approximation is essentially equivalent to the problem of 2 two-level atoms sharing a single excitation under a rotating wave and an essential states approximation. Thus, vacuum-induced coherence in the degenerate three-level vee-type system is directly analogous to the super- and sub-radiant effects in degenerate two-level atoms first emphasized by Dicke \cite{Dicke1954}.

In the Markov limit, when $\gamma\gg\omega_0,\Omega,\kappa^2$, we find that
\begin{equation}
  Q(t)\simeq\frac{\kappa^2}{2},
\end{equation}
so that the master equation in the Markov limit is given by
\begin{align}
  \partial_t\rho_t=-\frac{i}{\hbar}[\hsys,\rho_t]+\frac{\Gamma_1+\Gamma_2}{2}\left([\leff\rho_t,\leff^\dagger]+[\leff,\rho_t\leff^\dagger]\right),\label{masterveemarkov}
\end{align}
as can be verified using traditional methods \cite{Breuer2006}. We see in \eqref{masterveemarkov} that the decay rate out of $\ket{\phi^+}$ is the sum of the individual decay rates out of levels 1 and 2.

\begin{figure}
  \centering
    \includegraphics[width=\columnwidth]{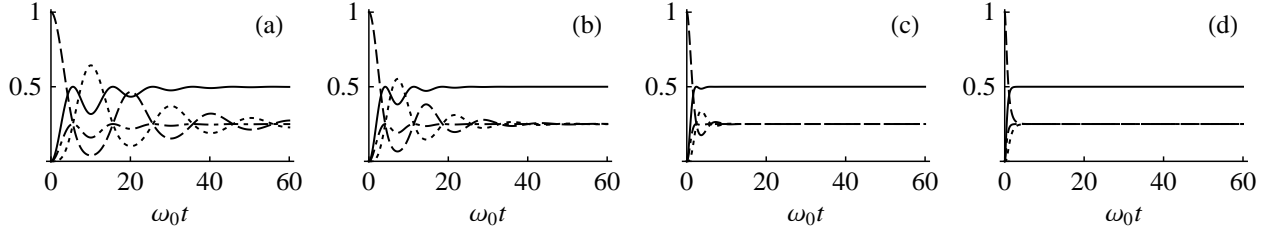}
    \caption{(a-d) Plot of $\rho_{33}$ (solid), $\rho_{11}$ (dashed), $\rho_{22}$ (dotted), and $|\rho_ {12}|$ (dot-dashed) when $\Gamma_m=\omega_0$, $\Delta=\omega_0/100$, and $\gamma_m=\gamma=\{0.1,0.2,1,5\}\omega_0$ for the initial state $\rho(0)=\proj{1}$. In all plots the system state in the long-time limit is given by $\rho_{11}=\rho_{22}=|\rho_{12}|=0.25$ and $\rho_{33}=0.5$.} 
    \label{plot1}
  \end{figure}

We now consider the case where the decay rates are equal $\Gamma_m=\Gamma$. Under this condition, the density matrix equations in the Markov limit \eqref{masterveemarkov} are exactly those found by Agarwal \cite{Agarwal1974}. We plot the evolution of the level populations, $\rho_{11}$, $\rho_{22}$, $\rho_{33}$, and the excited state coherence $|\rho_ {12}|$ in Fig.~\ref{plot1}a-d for various choices of the width of the coupling coefficients, $\gamma$, for the initial state $\rho(0)=\proj{1}$. 

We choose the coupling to be very strong, $\Gamma=\omega_0$, so that the decoherence evolution takes place in a few hundred cycles as opposed to many thousand cycles when $\Gamma\ll\omega_0$. We also choose the central frequencies of the coupling coefficients to be nearly resonant with the excited state energy, $\Delta=(\omega_0-\Omega)=-\omega_0/100$. As the width of the coupling coefficient increases from $\gamma=0.1\omega_0$ we see that the highly non-Markovian evolution in Fig.~\ref{plot1}a is replaced by the Markovian decay in Fig.~\ref{plot1}d where $\gamma=5\omega_0$. We also see that after a long time the total system state evolves to a mixture of $\ket{\phi^-}$ and the ground state,
\begin{equation}
  \rho_t\rightarrow\frac{1}{2}\proj{\phi^-}+\frac{1}{2}\proj{3},
\end{equation}
but that the time required to arrive at the long-time limit depends upon the correlation time, $1/\gamma$, of the reservoir operator $B$. 

\section{Multi-channel coupling with variable widths}
\begin{figure}
    \includegraphics[width=\columnwidth]{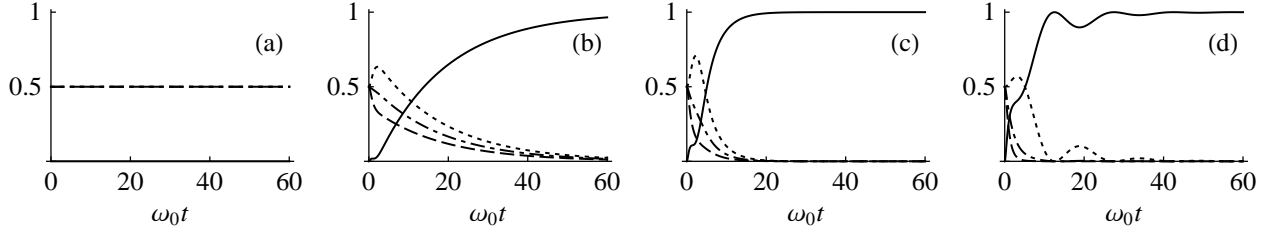}
    \caption{(a-d) Plot of $\rho_{33}$ (solid), $\rho_{11}$ (dashed), $\rho_{22}$ (dotted), and $|\rho_ {12}|$ (dot-dashed) when $\Gamma_m=\omega_0$, $\Delta=\omega_0/100$, $\gamma_1=5\omega_0$, and $\gamma_2=\{5,2.5,1,0.1\}\omega_0$ for the initial state $\rho(0)=\proj{\phi^-}$. In (a), $\rho_{22}=\rho_{11}=|\rho_{12}|=0.5$ as indicated by the overlap of the dashed, dotted, and dot-dashed lines.}
    \label{plot2}
\end{figure}

We now consider the case where the widths of the correlation functions begin to deviate from one another, but all other parameters are held fixed. When the widths of the coupling coefficients are not equal, the reservoir coupling is no longer to a single channel.  We investigate the case where the width of the coupling to level $1$ is given by $\gamma_1=5\omega_0$ and the initial state of the atom is $\ket{\phi^-}$ to examine how multi-channel coupling affects the evolution of the decoherence free state. In Fig.~\ref{plot2}a-d we plot the level populations $\rho_{mm}$, and the coherence between excited states $|\rho_{12}|$ as the width of the second coefficient, $\gamma_2$, decreases from the width of the first coefficient. As already discussed, the initial state satisfies the conditions of a decoherence free subspace when $\gamma_1=\gamma_2$, as shown in Fig.~\ref{plot2}a. As the width of the second coefficient decreases from the first, the system begins to decay to the ground state. When the width of the second coupling coefficient becomes very narrow relative to the first, $\gamma_2=0.1\gamma_1$, the decay of $\rho_{11}$ roughly reduces to what it would be if the decay from level 2 was forbidden, $\Gamma_2=0$. Consequently, we see that a common noise source mitigates the decay from level 1, even when a decoherence free subspace is not supported by the interaction.

\section{Multi-channel coupling with variable central frequencies }
\begin{figure}
  \centering
    \includegraphics[width=.75\columnwidth]{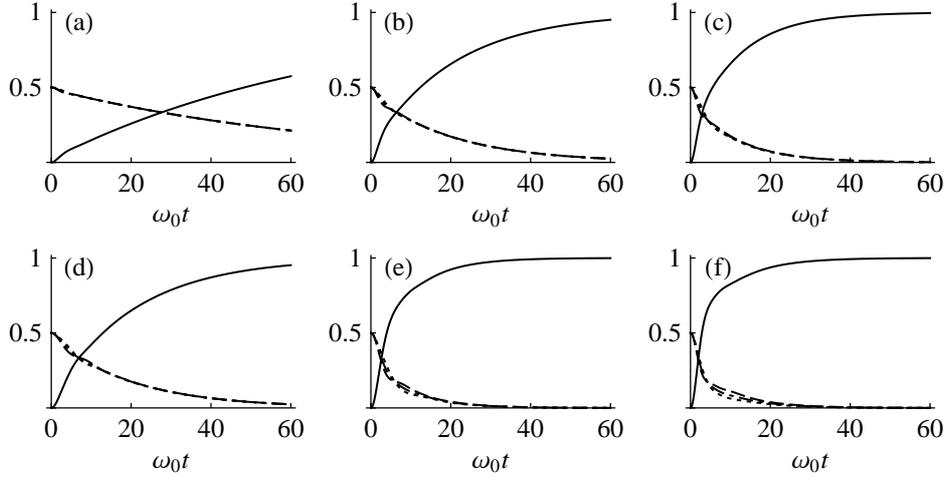}
    \caption{(a-f) Plot of $\rho_{33}$ (solid), $\rho_{11}$ (dashed), $\rho_{22}$ (dotted), and $|\rho_ {12}|$ (dot-dashed) when $\Gamma=\omega_0$, and $\Omega_1=1.01\omega_0$ for the initial state $\rho(0)=\proj{\phi^-}$. Plots (a-c) in the left column have $\gamma=\omega_0$ whereas plots (d-f) in the right column have $\gamma=0.5\omega_0$. The plots in the top (a,d), middle (b,e), and bottom (c,f) rows have $\Omega_2=\{.67,.33,0\}\omega_0$, respectively. In all plots, $\rho{11}$, $\rho_{22}$, and $|\rho_{12}|$ decay in nearly the same fashion from $0.5$ as indicated by the overlap of the dashed, dotted, and dot-dashed lines. }
    \label{plot3}
\end{figure}

Finally, we investigate the case where the width of the coupling coefficients are equal $\gamma_m=\gamma$, but the central frequencies are allowed to shift. When the central frequencies are not identical, the reservoir coupling is no longer to a single channel. We examine the case where the central frequency of the second coupling coefficient, $\Omega_2$, decreases from the first which remains near resonance, $\Omega_1=1.01\omega_0$, and plot the results in Fig.~\ref{plot3}a-f. Since the shift of the coupling coefficient relative to the width will be important, we investigate two cases: In Fig.~\ref{plot3}a-c we let $\gamma=\omega_0$ and let $\Omega_2=\{.67,.33,0\}\omega_0$, and in Fig.~\ref{plot3}d-f, we halve the bandwidth, $\gamma=0.5\omega_0$ and let the central frequency have the same values as before, $\Omega_2=\{.67,.33,0\}\omega_0$. 

Regardless of the width of the reservoir coupling, when central frequencies are the same, the initial state $\ket{\phi^-}$ does not decay, as shown for $\gamma=5\omega_0$ in Fig.~\ref{plot2}a. As the central frequency of the second channel deviates from the first, the frequency separation relative to width becomes important. In Fig.~\ref{plot3}a the frequency separation $|\Omega_2-\Omega_1|$ is only a third of the width. In contrast, Fig.~\ref{plot3}d has a frequency separation which is $2/3$ of the width. Consequently, the erosion of the decoherence free state in Fig.~\ref{plot3}d is more severe than in Fig.~\ref{plot3}a. This property is also apparent in comparing the rate of decay in Fig.~\ref{plot3}b and e, and c and f.

When the separation of the central frequencies is greater than their width, increasing the separation does not have a strong additional effect on the system evolution. This can be seen in Fig.~\ref{plot3}d-f; the rate of decay in Fig.~\ref{plot3}e and f remains roughly the same even though the frequency separation has increased in Fig.~\ref{plot3}f by the same amount as between Fig.~\ref{plot3}d and e. 

Figs.~\ref{plot1}-\ref{plot3} are also consistent with the result that the decoherence free subspaces are robust. Bacon \etal \cite{Bacon1999} showed that perturbations to interactions which support decoherence free subspaces do not degrade the decoherence free subspace to first order in the perturbation and to first order in time. To see how this applies to the plots above, we rewrite the interaction Hamiltonian as
\begin{align}
  \hint=\hint^++\hint^-
\end{align}
where $\hint^\pm=(\ket{3}\bra{\phi^\pm}\otimes B^\dagger_\pm+\hc)$, and where $B_\pm\propto B_1\pm B_2$. When $B_1\simeq B_2$, then $B_-\simeq0$, and we can consider $\hint^-$ as a perturbation to $\hint^+$. Consequently, to the degree that $B_1\simeq B_2$ we should expect similarly reduced decay out of $\ket{\phi_-}$ which is a decoherence free subspace of $\hint^+$. This is verified in Fig.~\ref{plot2}b, and Fig.~\ref{plot3}a, b, and d.  

\section{Conclusion}\label{conclusion}

We have extended the non-Markovian quantum state diffusion equation to open quantum systems with multi-channel reservoir coupling. Multi-channel reservoir coupling occurs when a canonical transformation of reservoir modes cannot reduce the number of reservoir operators appearing in the interaction Hamiltonian to one. Apart from fundamental interest in solving the most general type of reservoir coupling, open quantum systems which can exhibit multi-channel reservoir coupling are now under both theoretical and experimental investigation \cite{Lastra2011a}. Additionally, understanding the effect that multi-channel coupling has on decoherence will likely be important for developing practical quantum information systems.

For open quantum systems with multi-channel reservoir coupling which admit a noise-free O-operator, we have derived the exact master equation. We then considered non-Markovian evolution in the three-level vee-type system, finding the exact master equation in terms of time-dependent coefficients. By reformulating the master equation in terms of a time-dependent decay operator, we found the analytical solution to the general master equation for the three-level vee-type system with multi-channel coupling to a harmonic oscillator reservoir.

Using generalized Ornstein-Uhlenbeck noise we demonstrated how the solution to the  master equation for the three-level vee-type system may be used once the noise correlation function has been determined. When the correlations are identical (up to a constant), the multi-channel reservoir coupling reduces to single-channel reservoir coupling, and the coefficient equations \eqref{OUconsistency} can been solved analytically. In this limit, the decoherence free state corresponding to vacuum-induced coherence is supported by the interaction. As the spectral widths of the Ornstein-Uhlenbeck correlations deviate from each other, or as their central frequencies separate, the decoherence free state decays. When the correlations deviate strongly, in width or in central frequency, the decay proceeds much more rapidly than when the correlations are nearly the same. This demonstrates the robustness of decoherence free subspaces for Ornstein-Uhlenbeck noise correlations.

\section*{Acknowledgements}
JJ and TY acknowledge support from NSF PHY-0925174. CJB and JHE acknowledge support from ARO W911NF-09-1-0385 and NSF PHY-0855701.


\input{qsdbib.bbl}
\end{document}

%% file: qsd_header.tex
\def\hc{\textup{h.c.}}

\def\ak{a_k}

\def\ok{\omega_k}

\def\hb{\hbar}

\newcommand{\traceq}[1]{\textup{tr}\left[ #1 \right]}

\def\schrodinger{Schr\"{o}dinger }

\newcommand{\ket}[1]{|#1\rangle}                    
\newcommand{\bra}[1]{\langle #1|}                   

\newcommand{\braket}[2]{\left< #1 \vphantom{#2} \right|\left.\! #2 \vphantom{#1} \right>} 

\newcommand{\oprod}[2]{\ket{#1}\bra{#2}}
\newcommand{\proj}[1]{\oprod{#1}{#1}}